\documentclass[11pt]{article}
\usepackage{amssymb,amsmath,epsf,graphicx}
\input epsf.sty
\newcommand{\be}{\begin{equation}}
\newcommand{\ee}{\end{equation}}
\newcommand{\bea}{\begin{eqnarray}\displaystyle}
\newcommand{\eea}{\end{eqnarray}}
\newcommand{\bdm}{\begin{displaymath}}
\newcommand{\edm}{\end{displaymath}}
\newcommand{\sectiono}[1]{\section{#1}\setcounter{equation}{0}}

\newcommand{\starp}{*}

\newcommand{\bpz}{\star}

\def\ket#1{|#1 \rangle}

\def \ll {{\cal L}}

\def \bb {{\cal B}}

\begin{document}
{}~ \hfill\vbox{\hbox{hep-th/0606142}\hbox{MAD-TH-06-6}\hbox{CERN-PH-TH/2006-114} }\break
\vskip 2.1cm

\centerline{\Large \bf Proof of vanishing cohomology at the tachyon vacuum } \vspace*{2.0ex}
\vspace*{8.0ex}

\centerline{\large \rm Ian Ellwood$^a$ and Martin Schnabl$^b$}

\vspace*{8.0ex}

\centerline{\large \it $^a$Department of Physics, }
\centerline{\large \it University of Wisconsin, Madison, WI 53706, USA} \vspace*{2.0ex}
\centerline{E-mail: {\tt iellwood@physics.wisc.edu}}

\vspace*{6.0ex}

\centerline{\large \it $^b$Department of Physics, Theory Division,}
\centerline{\large \it CERN, CH-1211, Geneva 23, Switzerland} \vspace*{2.0ex}
\centerline{E-mail: {\tt martin.schnabl@cern.ch}}

\vspace*{6.0ex}

\centerline{\bf Abstract}
\bigskip

We prove Sen's third conjecture that there are no on-shell
perturbative excitations of the tachyon vacuum in open bosonic string
field theory.  The proof relies on the existence of a special state
$A$, which, when acted on by the BRST operator at the tachyon vacuum,
gives the identity.  While this state was found numerically in
Feynman-Siegel gauge, here we give a simple analytic expression.

\vfill \eject

\baselineskip=16pt

\newpage

\sectiono{Introduction}
\label{s_intro}

Following Sen's famous three conjectures \cite{Sen:1999mh,Sen:1999xm},
there has been an intensive effort to study the physics of tachyon
condensation in Witten's cubic open string field theory
\cite{Witten:1985cc}.  The power of open string field theory (OSFT)
over conventional CFT methods is that OSFT is an off-shell formulation
of open string interactions.  Many questions about open string vacua,
which must be understood using indirect arguments in CFT, can be
rephrased in OSFT as questions about the classical solutions of the
OSFT equations of motion.

Unfortunately, finding solutions to the OSFT equations of motion is
non-trivial.  Indeed, in the standard oscillator basis, these equations
become an infinite number of coupled non-linear differential equations
and, until recently, much of the work in OSFT has been numerical.

In spite of the approximate nature of the analysis, it has been found that
OSFT has a rich structure.  Starting from perturbative vacuum on the
D25-brane, one can find classical solutions to the equations of motion
representing lower-dimensional branes
\cite{Moeller:2000jy,deMelloKoch:2000xf,Moeller:2000hy,Beccaria:2005js} as well as the
tachyon vacuum
\cite{Kostelecky:1989nt,Sen:1999nx,Moeller:2000xv,Gaiotto:2002wy}, in
which there are no branes present.  In each case, the energy of these
solutions precisely matches the energy of the relevant brane
configuration, beautifully demonstrating Sen's first and second
conjectures.

Having found solutions representing various vacua, one can attempt to
find the spectrum of perturbative states around each solution.  In
particular, Sen's third conjecture states that around the tachyon
vacuum, which represents the absence of any brane at all, there should
be no physical states.  This conjecture has been checked in two
complementary ways.  First, the kinetic terms and gauge
transformations of certain low-mass excitations were computed to verify
that, indeed, there were no on-shell states
\cite{Ellwood:2001py,Giusto:2003wc}.  Second, it was argued that the
full spectrum of states was empty using a trick, which we now describe
\cite{Ellwood:2001ig}.

The physical states around a given vacuum are given by the cohomology
of a BRST operator $Q_\Psi$.  It turns out that the cohomology of
$Q_\Psi$ vanishes -- meaning that there are no physical states -- if
and only if there exists a state $A$ such that $Q_\Psi A =
\mathcal{I}$, where $\mathcal{I}$ is the identity of the star algebra.
Hence, the problem of showing that $Q_\Psi$ has vanishing cohomology
reduces to determining whether there is a solution to a single linear
equation.  This makes the problem amenable to numerical analysis and
it was found in \cite{Ellwood:2001ig} that, within the
level-truncation approximation, one could find such a state $A$.

Recently, one of us found an analytic solution to the OSFT equations
of motion representing the tachyon vacuum \cite{Schnabl:2005gv}.  This
solution has now been checked to satisfy the equations of motion, even
when contracted with itself \cite{Okawa:2006vm,Fuchs:2006hw}, and has
the correct energy \cite{Schnabl:2005gv}, giving an analytic proof of
Sen's first conjecture.  This solution opens up the possibility that
other questions in OSFT, which previously had only been understood
numerically, may have nice analytic solutions.

Indeed, in this paper we give a simple proof of Sen's third
conjecture.  We do this following the method described above: Given
the analytic solution $\Psi$, we find an analytic expression for a
state $A$ that satisfies $Q_\Psi A = \mathcal{I}$.

The organization of this paper is as follows: In section
\ref{s:ReviewOfOSFT}, we review the relevant aspects of OSFT.  In
section \ref{s:TachyonVacuum} we present the recently found analytic
solution to the equations of motion, $\Psi$.  Next, in section
\ref{s:VanishingCohomologyProof}, we define a new string field $A$,
which we then prove satisfies $Q_\Psi A = \mathcal{I}$. Finally, in
section \ref{s:pg}, we discuss the fact that the tachyon vacuum is a limit
of a family of pure-gauge solutions and show how this does not spoil
our cohomology arguments.



\sectiono{Review of OSFT}
\label{s:ReviewOfOSFT}

We begin with a review of some basic aspects of Witten's cubic open
string field theory.  Since there are many excellent reviews of OSFT
\cite{Thorn:1988hm,Sen:1999nx,Zwiebach:2001nj,Taylor:2003gn}, we will
only touch on some of the more relevant points.  The action is given
by \cite{Witten:1985cc}
\begin{equation} \label{WittenAction}
  S = \frac{1}{2} \int \Phi \starp Q_B \Phi + \frac{1}{3} \int \Phi
  \starp \Phi \starp \Phi.
\end{equation}
The classical field, $\Phi$, is an element of the free string Fock
space.  For example, for OSFT on a D25-brane background, it has an
expansion,
\begin{equation}
  \Phi = \int dp \left\{ t(p) + A_\mu (p) \alpha_{-1}^\mu + \psi(p)
  c_0 + \ldots \right\} c_1 |p\rangle,
\end{equation}
where $t(p)$ is the tachyon, $A_\mu(p)$ is the gauge field and $\psi(p)$ is a
ghost field.

The action (\ref{WittenAction}) has a large gauge invariance, which
makes solving the equations of motion in the non-gauge-fixed theory
difficult\footnote{The commutator is taken using the star product and
is graded by ghost number.  Explicitly,
\[
[\Phi_1 , \Phi_2] = \Phi_1
\starp \Phi_2 - (-1)^{\text{gh}(\Phi_1) \text{gh}(\Phi_2)} \Phi_2 \starp
\Phi_1
\]
};
\begin{equation}
  \Phi \to \Phi + Q_B \Lambda + [\Phi, \Lambda].
\end{equation}
Globally, fixing a gauge is a subtle issue
\cite{Ellwood:2001ne}. Around the perturbative vacuum, however, a
suitable choice is Feynman-Siegel gauge;
\begin{equation}
  b_0 \Phi = 0.
\end{equation}
Most of the numerical work in OSFT was performed in this gauge.
However, as we will discuss shortly, there is a different gauge which
is more suitable for analytic analysis.

The equations of motion of (\ref{WittenAction}) are given by
\begin{equation}
  Q_B \Psi + \Psi \starp \Psi = 0.
\end{equation}
Given a solution, $\Psi$, one can re-expand the action around the new vacuum;
\begin{equation}
  S(\Psi + \Phi) = \frac{1}{2} \int \Phi \starp Q_\Psi \Phi
  + \frac{1}{3} \int \Phi \starp \Phi \starp \Phi + \text{constant}.
\end{equation}
The new action takes the same form as the old action: the cubic term
is left completely invariant, while the kinetic term is only modified
by a change in the BRST operator, $Q_B \to Q_\Psi$, where
\begin{equation}
  Q_\Psi \Lambda = Q_B \Lambda + [\Psi,\Lambda].
\end{equation}
It is straightforward to check that $Q_\Psi^2 = 0$ using the equations
of motion of $\Psi$.  Just as the spectrum around the perturbative
vacuum was given by the cohomology of $Q_B$, the spectrum around the
new vacuum is given by the cohomology of $Q_\Psi$.

\subsection{OSFT in the $\arctan(z)$ coordinate system}

\begin{figure}
\begin{center}
\setlength{\unitlength}{1pt}
\begin{picture}(380,132)(0,0)%
\includegraphics{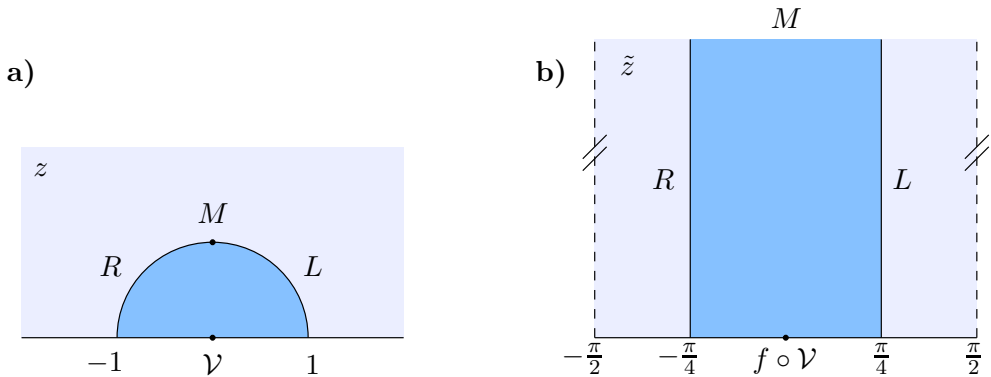}%
\end{picture}
\begin{picture}(0,0)(380,0)%
\put(10,75){\makebox(0,0)[lb]{$z$}}
\put(112,37){\makebox(0,0)[lb]{$L$}}
\put(35,37){\makebox(0,0)[lb]{$R$}}
\put(72,57){\makebox(0,0)[lb]{$M$}}
\put(232,113){\makebox(0,0)[lb]{$\tilde{z}$}}
\put(74,0){\makebox(0,0)[lb]{$\mathcal{V}$}}
\put(282,0){\makebox(0,0)[lb]{$f\circ\mathcal{V}$}}
\put(244,70){\makebox(0,0)[lb]{$R$}}
\put(335,70){\makebox(0,0)[lb]{$L$}}
\put(30,0){\makebox(0,0)[lb]{$-1$}}
\put(113,0){\makebox(0,0)[lb]{$1$}}
\put(210,0){\makebox(0,0)[lb]{$-\frac{\pi}{2}$}}
\put(246,0){\makebox(0,0)[lb]{$-\frac{\pi}{4}$}}
\put(327,0){\makebox(0,0)[lb]{$\frac{\pi}{4}$}}
\put(362,0){\makebox(0,0)[lb]{$\frac{\pi}{2}$}}
\put(0,108){\makebox(0,0)[lb]{{\bf a)}}}
\put(200,108){\makebox(0,0)[lb]{{\bf{b)}}}}
\put(289,131){\makebox(0,0)[lb]{$M$}}
\end{picture}
\end{center}
\caption{ \small The string field as seen by two coordinate systems.
In a) the standard description on upper half plane is illustrated.  A
vertex operator $\mathcal{V}$ generates a state on the unit
circle. The right half, left half and midpoint of the string are
labeled as viewed from infinity.  Diagram b) gives the same state in
the $\tilde{z} = \arctan(z)$ coordinate.  The left and right sides of
the figure are identified to give a cylinder. The left/right half of
the string now lies along the line $\Re(\tilde{z}) = \pm \pi/4$.  The
midpoint of the string is mapped to infinity.}
\label{f:arctanmap}
\end{figure}

Most of the difficulty in working with OSFT arises from the complexity
of the star product.  It was one of the key realizations of
\cite{Rastelli:2001vb,Schnabl:2005gv}, however, that the star product
simplifies when written in a different coordinate frame.

The standard method for specifying states in open string theory is by
putting a vertex operator, $\mathcal{V}$, on the boundary of the upper
half plane at the point $z= 0$.  By the operator-state correspondence
we can associate with $\mathcal{V}$ a state $|\mathcal{V}\rangle$ in
the string Fock space that lives on the unit circle.

\begin{figure}
\begin{center}
\setlength{\unitlength}{1pt}
\begin{picture}(235,132)(0,0)%
\includegraphics{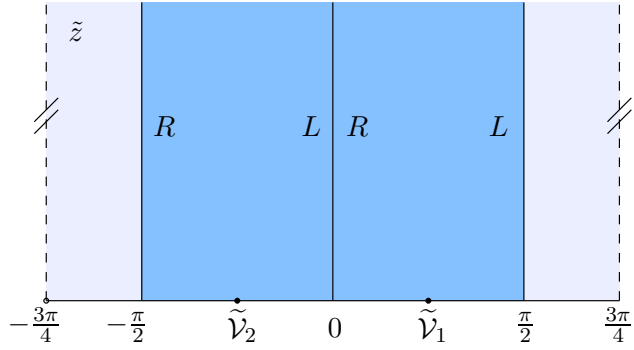}%
\end{picture}
\begin{picture}(0,0)(235,0)%
\put(14,112){\makebox(0,0)[lb]{$\tilde{z}$}}
\put(-9,-2){\makebox(0,0)[lb]{$-\frac{3\pi}{4}$}}
\put(28,-2){\makebox(0,0)[lb]{$-\frac{\pi}{2}$}}
\put(112,0){\makebox(0,0)[lb]{$0$}}
\put(183,-2){\makebox(0,0)[lb]{$\frac{\pi}{2}$}}
\put(216,-2){\makebox(0,0)[lb]{$\frac{3\pi}{4}$}}
\put(74,-2){\makebox(0,0)[lb]{$\widetilde{\mathcal{V}}_2$}}
\put(146,-2){\makebox(0,0)[lb]{$\widetilde{\mathcal{V}}_1$}}
\put(119,76){\makebox(0,0)[lb]{$R$}}
\put(46,76){\makebox(0,0)[lb]{$R$}}
\put(102,76){\makebox(0,0)[lb]{$L$}}
\put(173,76){\makebox(0,0)[lb]{$L$}}
\end{picture}
\end{center}
\caption{ \small A pictorial description of the star product.  Given
two states $|\tilde{\mathcal{V}}_1\rangle $ and
$|\tilde{\mathcal{V}}_2\rangle$ generated by inserting vertex
operators $\tilde{\mathcal{V}}_1$ and $\tilde{\mathcal{V}}_2$ in the
$\tilde{z}$ coordinate, the star product,
$|\tilde{\mathcal{V}}_1\rangle \starp |\tilde{\mathcal{V}}_2\rangle$,
is computed by gluing the right side of the
$|\tilde{\mathcal{V}}_1\rangle $ state to the left side of the
$|\tilde{\mathcal{V}}_2\rangle$ state.  This gives a cylinder of width
$3 \pi/2$.  }
\label{f:starproduct}
\end{figure}

However, there was no reason why we had to choose the upper half plane
to define our states.  It turns out to be useful to work instead in
the coordinate $\tilde{z} = f(z) = \arctan(z)$.  Under $z \to f(z)$,
the upper half plane is mapped to an infinitely tall cylinder as
illustrated in figure~\ref{f:arctanmap}.  In this frame, the star
product can be described purely geometrically; one simply glues the
strips of world-sheet together that correspond to the two string
states.  This is illustrated in figure~\ref{f:starproduct}.  

Multiplying $n$ strips of width $\pi/2$ will produce a strip of width
$n \pi/2$ and it is useful to consider the class of all such states.
When there are no operator insertions, a state described by a strip of
width $n \pi/ 2$ is called a wedge state and is denoted $|n+1\rangle$.
These states were first introduced in \cite{Rastelli:2000iu}, and obey
the algebra,
\begin{equation}
  |n\rangle \starp |m\rangle = |m+n -1\rangle.
\end{equation}
The state $|2\rangle$ is just the original strip of width $\pi/2$ with
no vertex operator inserted at the origin and is, thus, the
$SL(2,\mathbb{R})$ invariant vacuum $|0\rangle$.  

It turns out that taking the limit as the width of the strip tends to
infinity leads to a finite state; $|\infty\rangle = \lim_{n\to \infty}
|n\rangle$.  This state is known as the sliver \cite{Rastelli:2000iu} and as is a projector
under star multiplication;
\begin{equation}
   |\infty \rangle \starp |\infty \rangle = |\infty \rangle.
\end{equation}
Notice that multiplying a state $\Lambda$ by the wedge state of zero width,
$|1\rangle$, leaves $\Lambda$ invariant.  Hence, $\mathcal{I} =
|1\rangle$ is an identity of the star algebra;
\begin{equation}
  \Lambda \starp \mathcal{I} = \mathcal{I} \starp \Lambda = \Lambda.
\end{equation}
A useful property of $\mathcal{I}$ is that, at least formally, for any operator $\mathcal{O}$ it
obeys \cite{Gross:1986ia,Gross:1986fk}
\begin{equation}\label{eq:OIequalsOIdagger}
  \mathcal{O} |\mathcal{I}\rangle  = \mathcal{O}^\bpz |\mathcal{I}\rangle
   = \tfrac{1}{2} ( \mathcal{O}+ \mathcal{O}^\bpz) |\mathcal{I}\rangle ,
\end{equation}
where, in the notation of \cite{RZ2006}, $\mathcal{O}^\bpz$ denotes
BPZ conjugation; $\mathcal{O}^\bpz = I\circ \mathcal{O}$, where $I(z)
= -1/z$.

\subsection{Some important operators}

In general, each of the familiar operators in the $\tilde{z}$
coordinate can be pulled back into the $z$ coordinate using
$f^{-1}(\tilde{z}) = \tan(\tilde{z})$.  We will occasionally denote
such an operator using a tilde; e.g. $\tilde{c}(\tilde{z}) = f^{-1}
\circ c(z)$. It is also useful to make the following definitions:
\begin{equation}
\qquad
  \mathcal{L}_0 = f^{-1} \circ L_0,
\qquad
  \mathcal{B}_0 = f^{-1} \circ b_0,
\qquad
  K_1 = f^{-1} \circ L_{-1},
\qquad
  B_1 = f^{-1} \circ b_{-1}.
\end{equation}
Just as $L_0$ gave the mass level of fields in the $z$ coordinate,
$\mathcal{L}_0$ is the analogous level in the $\tilde{z}$ coordinates.
Similarly, while the standard gauge fixing condition in the
$z$-coordinate was $b_0 \Phi = 0$, in the $\tilde{z}$-coordinate, one
uses $\mathcal{B}_0 \Phi = 0$.

Explicit mode expansions of $\mathcal{L}_0$ and  $\mathcal{B}_0$ are given by
\begin{align}
  \mathcal{L}_0 &= L_0 + \tfrac{2}{3} L_2 - \tfrac{2}{15} L_4 +\cdots,
\\
  \mathcal{B}_0 &= b_0 + \tfrac{2}{3} b_2 - \tfrac{2}{15} b_4 +\cdots.
\end{align}
Note that while $L_0$ and $b_0$ are BPZ dual to themselves, their
script cousins are not and we also have operators $\mathcal{L}_0^\bpz$
and $\mathcal{B}_0^\bpz$, which are given by $L_{n}^\bpz = (-1)^n
L_{-n}$ and $b_n^\bpz = (-1)^n b_{-n}$.  These obey the commutation
relations\footnote{These commutation relations are an important
property of the conformal frame of the sliver.  Recently
\cite{RZ2006}, it has been shown that the conformal frames of other
projectors, known as special projectors, lead to similar algebras;
$[\mathcal{L}_0,\mathcal{L}_0^\bpz] =
s(\mathcal{L}_0+\mathcal{L}_0^\bpz)$. These special projectors have
many similarities with the sliver and can be used to solve the
ghostnumber zero equations of motion \cite{Gaiotto:2002uk,RZ2006}.},
\begin{equation}\label{eq:LLandBBcommutators}
  [\mathcal{L}_0,\mathcal{L}_0^\bpz] = \mathcal{L}_0+\mathcal{L}_0^\bpz,
\end{equation}
as well as
\begin{equation} \label{eq:BLcommutators}
  [\mathcal{L}_0, \mathcal{B}_0] = [\mathcal{L}_0^\bpz, \mathcal{B}_0^\bpz]= 0,
\qquad
  [\mathcal{L}_0^\bpz, \mathcal{B}_0] = -\mathcal{B}_0-\mathcal{B}_0^\bpz,
\qquad
  [\mathcal{L}_0, \mathcal{B}_0^\bpz] = \mathcal{B}_0+\mathcal{B}_0^\bpz.
\end{equation}
An important property of $\mathcal{L}_0$ is that the wedge states can be represented in the form \cite{LeClair:1988sp,LeClair:1988sj,Rastelli:2000iu,Schnabl:2002gg}
\begin{equation} \label{eq:wedgeInTermsOfU}
  |r\rangle = U_r^\bpz |0\rangle,
\end{equation}
where $U_r = (2/r)^{\mathcal{L}_0}$.  The operators $U_r$ and $U_s^\bpz$ obey the
important relation \cite{Schnabl:2002gg},
\begin{equation} \label{eq:UUdaggerrelation}
  U_r U_s^\bpz = U_{2+ \frac{2}{r}(s-2)}^\bpz U_{2+\frac{2}{s} (r-2)},
\end{equation}
which can be used to derive (\ref{eq:LLandBBcommutators}).

The operators $K_1$ and $B_1$ take a very simply form,
\begin{equation}
  K_1 = L_1+L_{-1}, \qquad B_1 = b_1 + b_{-1}.
\end{equation}
These operators were first studied in \cite{Rastelli:2000iu}, where it
was shown that they are derivations of the star algebra;
\begin{align}\label{eq:K1B1asderivations}
  K_1 (\Phi_1 \starp \Phi_2) &=
     (K_1 \Phi_1) \starp \Phi_2 + \Phi_1 \starp (K_1 \Phi_2).
\\
  B_1 (\Phi_1 \starp \Phi_2) &=
     (B_1 \Phi_1) \starp \Phi_2 + (-1)^{\text{gh}(\Phi_1)} \Phi_1 \starp (B_1 \Phi_2).
\end{align}
They also annihilate the wedge states;
\begin{equation} \label{eq:LBLKonwedge}
  K_1|r\rangle =
  B_1|r\rangle =0.
\end{equation}
In the $\tilde{z}$ coordinates they take the form,
\begin{equation}
  K_1 = \oint d\tilde{z} \,T(\tilde{z}),
\qquad
  B_1 = \oint d\tilde{z} \,b(\tilde{z}).
\end{equation}
It is also useful to define the ``left'' and ``right'' parts of these
operators, which are given by taking only the left or right parts --
as viewed from infinity -- of the contour integral;
\begin{equation} \label{eq:KLRBLR}
   K_1^{L,R} = \oint_{\gamma^{L,R}} d\tilde{z}\,  \,T(\tilde{z}),
\qquad
   B_1^{L,R} = \oint_{\gamma^{L,R}} d\tilde{z}\,  \,b(\tilde{z}).
\end{equation}
In the $\tilde{z}$ coordinates, the contours, $\gamma^{L,R}$, are
given by the vertical lines on the right and left of the strip.  Note
that, $K_1^L + K_1^R = K_1$ and $B_1^L + B_1^R = B_1$. Also,
\begin{align}\label{eq:KLRonstarproduct}
  K_1^L (\Phi_1 \starp \Phi_2) &= (K_1^L\Phi_1 ) \starp \Phi_2,
&
  B_1^L (\Phi_1 \starp \Phi_2) &= (B_1^L\Phi_1 ) \starp \Phi_2,
\\
  K_1^R (\Phi_1 \starp \Phi_2) &= \Phi_1 \starp (K_1^R\Phi_2 ),
&
  B_1^R (\Phi_1 \starp \Phi_2) &= (-1)^{\text{gh}(\Phi_1)}\Phi_1 \starp (B_1^R\Phi_2 ).
\end{align}

An important property of the operators $K_1^{L,R}$ is that they act as
a derivative with respect to the width of the state.  This follows
from (\ref{eq:KLRBLR}). Since the $K^{L,R}$ are just integrals of $T$
in the $\tilde{z}$ coordinate and $\int T(\tilde{z})$ is the
world-sheet Hamiltonian, $\epsilon K^{R,L}$ can be thought of as
adding/subtracting an infinitesimal strip of with $\epsilon$ from the
right/left of the world-sheet.  This gives the useful identity,
\begin{equation} \label{eq:KLRonwedge}
   \partial_n |n\rangle = \pm \tfrac{\pi}{2} K_1^{R,L} |n\rangle,
\end{equation}
which can be integrated to give
\begin{equation} \label{eq:wedgefromKs}
  |n\rangle = e^{\pm \frac{\pi}{2} (n-2) K_1^{R,L}} |0\rangle.
\end{equation}

The operators $K_1^{L,R}$, $\mathcal{L}_0$ and $\mathcal{L}_0^\bpz$,
as well as $B_1^{L,R}$, $\mathcal{B}_0$ and $\mathcal{B}_0^\bpz$ are
related through the identities,
\begin{equation} \label{eq:KLRandLLdagger}
  K_1^L - K_1^R = \tfrac{2}{\pi} (\mathcal{L}_0 + \mathcal{L}_0^\bpz),
\qquad
  B_1^L - B_1^R = \tfrac{2}{\pi} (\mathcal{B}_0 + \mathcal{B}_0^\bpz),
\end{equation}
which follow from the definitions of these operators.  Using (\ref{eq:KLRandLLdagger}), we can rewrite (\ref{eq:wedgefromKs})
as
\begin{equation} \label{eq:wedgeasexpofLs}
  |n\rangle = e^{\frac{(2-n)}{2}
   (\mathcal{L}_0+\mathcal{L}_0^\bpz)} |0\rangle.
\end{equation}
This expression can be related to (\ref{eq:wedgeInTermsOfU}) using the identity,
\begin{equation}\label{eq:UUdagger}
    e^{\frac{(2-n)}{2}
   (\mathcal{L}_0+\mathcal{L}_0^\bpz)} = U_n^\bpz U_n.
\end{equation}
A more general collection of such identities can be found in
\cite{Schnabl:2002gg,Schnabl:2002ff,Schnabl:2002ef,Schnabl:2005gv}.

\section{The exact tachyon vacuum solution}
\label{s:TachyonVacuum}

In this section, we review the exact tachyon vacuum state found in
\cite{Schnabl:2005gv}.  Define
\begin{equation}
  \psi_n = \tfrac{2}{\pi} c_1 |0\rangle \starp B_1^L |n\rangle \starp
c_1| 0\rangle.
\end{equation}
Then the tachyon vacuum is given by\footnote{The term $-\partial_n
\psi_n$ for $n = 0$ can be defined by carefully taking the limit.
Explicitly, one finds $Q_B B_1^L c_1|0\rangle$.}
\begin{equation}
  \Psi =
\lim_{N\to \infty}
\left( \psi_N - \sum_{n=0}^N \partial_n \psi_n \right).
\end{equation}
Formally, the $\psi_N$ piece vanishes in level truncation as $N \to
\infty$, but it gives finite contributions to the energy and is
required for $\Psi$ to satisfy the equations of motion when contracted
with itself \cite{Okawa:2006vm,Fuchs:2006hw};
\begin{equation}
  \langle \Psi | Q_B \Psi\rangle +\langle \Psi |\Psi \starp \Psi \rangle  = 0.
\end{equation}
We will see that this term is also required to give a complete proof
that the cohomology of $Q_\Psi$ vanishes.

The solution satisfies the gauge fixing condition,
\begin{equation}
  \mathcal{B}_0 \Psi = 0,
\end{equation}
which as alluded to earlier, is the analogue of Feynman-Siegel gauge
fixing in the $\tilde{z}$-coordinate.

\begin{figure}
\begin{center}
\setlength{\unitlength}{1pt}
\begin{picture}(380,129)(0,0)%
\includegraphics{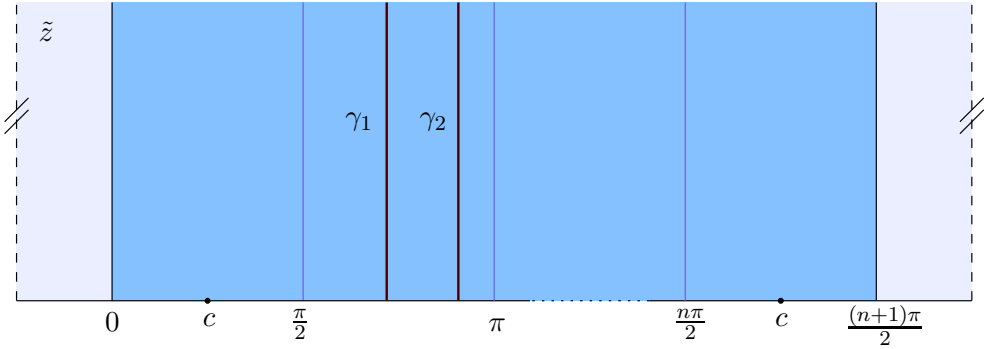}%
\end{picture}
\begin{picture}(0,0)(380,0)%
\put(14,112){\makebox(0,0)[lb]{$\tilde{z}$}}
\put(76,4){\makebox(0,0)[lb]{$c$}}
\put(292.5,4){\makebox(0,0)[lb]{$c$}}
\put(130,78){\makebox(0,0)[lb]{$\gamma_1$}}
\put(158,78){\makebox(0,0)[lb]{$\gamma_2$}}
\put(39,2){\makebox(0,0)[lb]{$0$}}
\put(109,-1){\makebox(0,0)[lb]{$\frac{\pi}{2}$}}
\put(184,2){\makebox(0,0)[lb]{$\pi$}}
\put(254,-1){\makebox(0,0)[lb]{$\frac{n \pi}{2}$}}
\put(319,-3){\makebox(0,0)[lb]{$\frac{(n+1) \pi}{2}$}}
\end{picture}
\end{center}
\caption{ \small The state $-\partial_n \psi_n$ is given by a strip of
width $\frac{\pi}{2} (n+1)$ with two insertions of $c(\tilde{z})$ as
well as two contour integrals of $T(\tilde{z})$ and $b(\tilde{z})$
along the curves $\gamma_1$ and $\gamma_2$.  }
\label{f:dpsi}
\end{figure}

The states $-\partial_n \psi_n$ can be written using
(\ref{eq:KLRonwedge}) as
\begin{equation}\label{eq:dpsidn}
  -\partial_n \psi_n =
c_1|0\rangle \starp B_1^L K_1^L |n\rangle \starp c_1 |0\rangle.
\end{equation}
These states take a simple form in the $\tilde{z}$ coordinate, as illustrated
in figure \ref{f:dpsi}.

\section{Proof that $Q_\Psi$ has no cohomology}
\label{s:VanishingCohomologyProof}

Having defined $\Psi$, we can now turn to the main aim of this paper,
to prove that $Q_\Psi$ has vanishing cohomology so that there are no
on-shell perturbative states around the tachyon vacuum.  As discussed
in the introduction, we can do this using a trick, which we state as a
simple lemma:

\bigskip

\noindent
{\em Lemma:} The cohomology of a BRST operator $Q_\Psi$ vanishes if
and only if there exists a string field $A$ such that $Q_\Psi A =
\mathcal{I}$.

\bigskip

\noindent
{\em Proof:} First, suppose that $Q_\Psi$ has no cohomology.  Consider
$Q_\Psi \mathcal{I} = Q_B \mathcal{I} + \Psi \starp \mathcal{I}
-\mathcal{I}\starp \Psi = Q_B \mathcal{I}$.  Since, as was first shown
in \cite{Gross:1986ia,Gross:1986fk}, $Q_B \mathcal{I} =0$, it follows
that $Q_\Psi \mathcal{I} = 0$.  Since $\mathcal{I}$ is $Q_\Psi$-closed
and $Q_\Psi$ has no cohomology, there must exist some $A$ such that
$\mathcal{I} = Q_\Psi A$.

Now suppose, instead, that we have a state $A$ such that $Q_\Psi A =
\mathcal{I}$.  Suppose we also have some $Q_\Psi$-closed state
$\Lambda$ such that $Q_\Psi \Lambda = 0$.  Then
\begin{equation}
  Q_\Psi (A\starp \Lambda) = (Q_\Psi A)\starp \Lambda = \mathcal{I} \starp \Lambda
 = \Lambda,
\end{equation}
so that $\Lambda$ is $Q_\Psi$-exact.  Since any $Q_\Psi$-closed state
is also $Q_\Psi$-exact, it follows that $Q_\Psi$ has no cohomology.

Such an operator $A$ is known in the math literature as a homotopy
operator.  Note that the existence of $A$ proves that the cohomology
of $Q_\Psi$ vanishes at all ghost numbers, not just ghost number zero
as required by Sen's conjectures.\footnote{This seems to contradict
the numerical results of \cite{Giusto:2003wc}.  Nonzero cohomology at
other ghost numbers has also been found for the so-called universal
solution \cite{Takahashi:2002ez} in \cite{Kishimoto:2002xi}.}

\subsection{Finding the state $A$}

We now describe how to find an $A$ satisfying,
\begin{equation}\label{eq:QPsiA}
  Q_\Psi A = Q_B A + \Psi \starp A + A \starp \Psi = \mathcal{I}.
\end{equation}
Although (\ref{eq:QPsiA}) is a linear equation for $A$, a blind search
for a solution could be very difficult.  Fortunately, for the
Feynman Siegel gauge solution, (\ref{eq:QPsiA}) was solved numerically
in \cite{Ellwood:2001ig} and we can use the results found there to guess a solution.

Surprisingly, it was found in \cite{Ellwood:2001ig} that, in Feynman-Siegel gauge,
$A$ takes the approximate form,
\begin{equation}
  A_{\text{FS}} \sim \frac{1}{L_0} b_0\mathcal{I}.
\end{equation}
Curiously, this form of $A_{\text{FS}}$ is the state one would write
down if one was trying to show that, in the {\em perturbative} vacuum,
$Q_B$ had vanishing cohomology.  Indeed one has
\begin{equation}\label{QBAFS}
  Q_B A_{\text{FS}} = \mathcal{I} - |0\rangle,
\end{equation}
so that one finds the identity state minus the one piece of the
identity that is in the cohomology of $Q_B$.

\begin{figure}
\begin{center}
\setlength{\unitlength}{1pt}
\begin{picture}(136,140)(0,0)%
\includegraphics{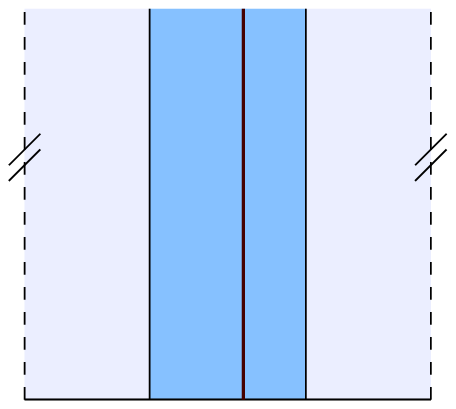}%
\end{picture}
\begin{picture}(0,0)(136,0)%
\put(12,118){\makebox(0,0)[lb]{$\tilde{z}$}}
\put(57,86){\makebox(0,0)[lb]{$\gamma$}}
\put(42,1){\makebox(0,0)[lb]{$\underbrace{\hspace{1.5cm}}_{\pi(r-1)/2}$}}
\put(-70,56){\makebox(0,0)[lb]{\parbox{1cm}{$$A =
\tfrac{\pi}{2} \int_1^2 dr$$}}}
\end{picture}
\end{center}
\caption{ \small The state $A$ can be represented as a sum over wedge
states $|r\rangle$. The only operator insertion is a single contour
integral of $b(\tilde{z})$ along the curve $\gamma$.  }
\label{f:Astate}
\end{figure}

A natural guess for the $\mathcal{B}_0$-gauge solution is to take the
same form for $A$, but with $b_0$ and $L_0$ replaced by their
counterparts in the $\tilde{z}$ coordinate, $\mathcal{B}_0$ and
$\mathcal{L}_0$;
\begin{equation}
  A = \frac{1}{\mathcal{L}_0} \mathcal{B}_0 \mathcal{I}.
\end{equation}
It turns out that this $A$ can be written in a nicer form, as an
integral over wedge states with insertions.  Using
(\ref{eq:OIequalsOIdagger}), we have
\begin{equation} \label{eq:BtoBBdagger}
  A = \frac{1}{2\mathcal{L}_0} (\mathcal{B}_0 +\mathcal{B}^\bpz_0)
  \mathcal{I}.
\end{equation}
Since $(\mathcal{B}_0 + \mathcal{B}_0^\bpz)$ raises the
$\mathcal{L}_0$-level by one, we may rewrite (\ref{eq:BtoBBdagger}) as
\begin{equation}
  A = \tfrac{1}{2} (\mathcal{B}_0 +\mathcal{B}^\bpz_0)
  \frac{1}{\mathcal{L}_0+1} \mathcal{I}.
\end{equation}
This can be further simplified by writing
\begin{equation}
  \frac{1}{\mathcal{L}_0+1}
   = \int_0^1  z^{\mathcal{L}_0} dz
   = \int_0^1 dz\, U_{2/z} .
\end{equation}
Using (\ref{eq:UUdaggerrelation}), we have
\begin{equation}
  U_{2/z} \mathcal{I} = U_{2/z} U_1^\bpz |0\rangle
   = U_{2-z}^\bpz  |0\rangle = |2-z\rangle,
\end{equation}
which yields\footnote{This result can also be found directly in the
$\mathcal{L}_0$-level expansion;
\begin{multline}
A = \frac{1}{2\mathcal{L}_0} (\mathcal{B}_0
+ \mathcal{B}_0^\bpz)
\sum_{n = 0}^\infty \frac{1}{2^n n!} (\mathcal{L}_0
+ \mathcal{L}_0^\bpz)^n |0\rangle
 = \frac{1}{2} \sum_{n = 0}^\infty (\mathcal{B}_0
+ \mathcal{B}_0^\bpz)
\frac{1}{2^n (n+1)!} (\mathcal{L}_0
+ \mathcal{L}_0^\bpz)^n |0\rangle
\\
= \tfrac{1}{2} (\mathcal{B}_0
+ \mathcal{B}_0^\bpz)
   \frac{
      e^{\frac{1}{2}
          (\mathcal{L}_0 + \mathcal{L}_0^\bpz)} - 1}{(\mathcal{L}_0
+ \mathcal{L}_0^\bpz)/2} |0\rangle
 = \tfrac{1}{2} (\mathcal{B}_0
+ \mathcal{B}_0^\bpz)
\int_1^2 dr\, e^{\frac{2-r}{2} (\mathcal{L}_0 + \mathcal{L}_0^\bpz)}
|0\rangle = \tfrac{1}{2}  (\mathcal{B}_0
+ \mathcal{B}_0^\bpz) \int_1^2 dr\, |r\rangle.
\end{multline}
}
\begin{equation}
  A = \tfrac{1}{2} (\mathcal{B}_0+\mathcal{B}_0^\bpz) \int_0^1 dz
   \,|2-z\rangle = \tfrac{1}{2} (\mathcal{B}_0+\mathcal{B}_0^\bpz)
   \int_1^2 dr\, |r\rangle.
\end{equation}
Using (\ref{eq:KLRandLLdagger}) and (\ref{eq:LBLKonwedge}) this becomes
\begin{equation}
  A = \tfrac{\pi}{2} B^{L}_1 \int_1^2 dr\, |r\rangle.
\end{equation}
This state has a simple geometric interpretation, as shown in
figure~\ref{f:Astate}.

\subsection{Computation of $Q_\Psi A$}

The first term in $Q_\Psi A$ is just $Q_B A$.  This is given by
\begin{equation}\label{eq:QBA}
  Q_B A = \tfrac{\pi}{2} K_1^L \int_1^2 |r\rangle
        = - \int_1^2 dr\, \partial_r |r\rangle
        = \mathcal{I} - |0\rangle,
\end{equation}
which reproduces the Feynman-Siegel gauge result, (\ref{QBAFS}).

Next we must compute the star-products $\Psi\starp A$ and $A\starp
\Psi$.  Because the tachyon vacuum solution is twist invariant, these
two computations are related to each other by a twist.  Hence, we need
to compute just one of them, $\Psi\starp A$.

Since $\Psi = \psi_N - \sum_{m = 0}^n \partial_n \psi_n$, we begin by
evaluating $ \psi_n\starp A$.  Using (\ref{eq:dpsidn}), we have
\begin{equation}
   \psi_n\starp A = \int_1^2 dr \, c_1|0\rangle \starp B_1^L |n\rangle
\starp c_1 |0\rangle
\starp B_1^L |r\rangle,
\end{equation}
which we can rewrite using (\ref{eq:KLRonstarproduct}) as
\begin{equation}
 \int_1^2 dr \, c_1|0\rangle \starp  B_1^L B_1^R(|n\rangle
\starp c_1 |0\rangle
\starp  |r\rangle).
\end{equation}
Now, using $B_1^L B_1^R = B_1^L (B_1-B_1^L) = B_1^L B_1$ this becomes
\begin{equation}
 \int_1^2 dr \, c_1|0\rangle \starp  B_1^L B_1(|n\rangle
\starp c_1 |0\rangle
\starp  |r\rangle)
=\int_1^2 dr \, c_1|0\rangle \starp  B_1^L (|n\rangle
\starp |0\rangle
\starp  |r\rangle),
\end{equation}
where we have used the derivation property
(\ref{eq:K1B1asderivations}) of $B_1$ as well as
(\ref{eq:LBLKonwedge}).  It follows that
\begin{equation} \label{eq:psistarA}
  \psi_n \starp A = \int_1^2 dr \, B_1^R(c_1|0\rangle \starp  |n+r\rangle).
\end{equation}
Similarly, one can compute $A\starp \psi_n$ either by repeating the
above computation or by exploiting twist symmetry.  Either way, one
finds
\begin{equation}\label{eq:Astarpsi}
  A\starp \psi_n = \int_1^2 dr\, B_1^L (|r+n\rangle \starp c_1 |0\rangle).
\end{equation}
Now consider $- \sum_{n=0}^N \partial_n \psi_n\starp A$.  Since $n$
and $r$ appear only in the combination $n+r$, we can replace the
derivative $\partial_n$ with $\partial_r$.  This gives
\begin{equation}
  - \sum_{n=0}^N \partial_n \psi_n\starp A = -\sum_{n=0}^N \int_1^2 dr
 \, B_1^R(c_1|0\rangle \starp \partial_r |n+r\rangle)
 = \sum_{n=0}^N
 \, B_1^R(c_1|0\rangle \starp \{|n+1\rangle - |n+2\rangle\}).
\end{equation}
Notice that the sum can now be trivially performed since
\begin{equation}
  \sum_{n = 0}^N |n+1\rangle - |n+2\rangle
 = \mathcal{I} - |N+2\rangle.
\end{equation}
Hence, we find
\begin{equation}\label{eq:dpsistarA}
  - \sum_{n=0}^N \partial_n \psi_n\starp A
 = B_1^R c_1|0\rangle
   - B_1^R (c_1|0\rangle \starp |N+2\rangle ).
\end{equation}
Similarly, one can compute
\begin{equation}\label{eq:Astardpsi}
  - A \starp \sum_{n=0}^N \partial_n \psi_n
=
  B_1^L c_1|0\rangle
   - B_1^L (|N+2\rangle \starp c_1|0\rangle  ).
\end{equation}
Using (\ref{eq:psistarA}), (\ref{eq:Astarpsi}), (\ref{eq:dpsistarA})
and (\ref{eq:Astardpsi}), we find, in total, that
\begin{equation}
  \Psi\starp A+ A\starp \Psi
 = |0\rangle -\Sigma,
\end{equation}
where the state $\Sigma$ is given by
\begin{equation}
\Sigma=
B_1^R\left\{c_1|0\rangle \starp
    \left(|N+2\rangle - \int_1^2 dr \,|N+r\rangle\right)\right\}
\\
   +B_1^L\left\{ \left(|N+2\rangle - \int_1^2 dr \,|N+r\rangle\right)
    \starp c_1|0\rangle\right\}.
\end{equation}
Now, as $N \to \infty$, the state $|N\rangle$ limits to the sliver so
that $|N+2\rangle - \int_1^2 dr\, |N+r\rangle \to 0$ as $N \to
\infty$. In fact, it is straightforward to check that, in the level
expansion, it goes to zero as $\mathcal{O}(N^{-3})$.  Hence, when we
remove the regulator we find
\begin{equation} \label{psistarAplusAstarpsi}
  \Psi \starp A + A \starp \Psi = |0\rangle.
\end{equation}
Note that it was important to include the $\psi_N$ piece in $\Psi$ to
cancel out the surface terms in the sums (\ref{eq:dpsistarA}) and
(\ref{eq:Astardpsi}).  Combining (\ref{psistarAplusAstarpsi}) with
(\ref{eq:QBA}), we find the desired result;
\begin{equation}
  Q_\Psi A = Q_B A + \Psi \starp A + A\starp \Psi = \mathcal{I}.
\end{equation}
This proves that the cohomology of $Q_\Psi$ is empty.

\subsection{Comparison with vacuum string field theory}
\label{s:Aproperties}
It is interesting to compare our results with the results of vacuum
string field theory (VSFT) \cite{Rastelli:2000hv,Rastelli:2001uv}.  In
VSFT, the BRST operator around the tachyon vacuum is taken, by ansatz,
to be a simple pure ghost operator.  For example, one of the early
choices was the zero mode of the $c$-ghost, $c_0$.  To show that $c_0$
has empty cohomology, one notes that $\{c_0,b_0\} = 1$, so that $b_0$
plays the role of our string field, $A$.

This analogy can be made a little closer. Just as $b_0^2 = 0$, it
happens that $A\starp A =0$. This property is easy to see from the
geometric form of $A$ in figure \ref{f:Astate}.  Moreover, just as
$b_0$ is a Hermitian operator, one can also construct a Hermitian
operator $\hat{A}$ defined by
\begin{equation}
  \hat{A} \Phi = A\starp \Phi + (-1)^{\text{gh}(\Phi)} \Phi \starp A,
\end{equation}
which satisfies $\hat{A}^2 = 0$ and $\{Q_\Psi ,\hat{A}\} = 1$, as well
as the Hermiticity property, $\langle \Phi_1 | \hat{A} \Phi_2\rangle =
\langle \hat{A} \Phi_1 | \Phi_2\rangle$.  Since VSFT is thought to be
a singular limit of ordinary OSFT, in which the BRST operator becomes a
$c$-ghost operator inserted at the midpoint \cite{Gaiotto:2001ji}, it
would be interesting to see whether $\hat{A}$ becomes a simple
operator formed out of just the $b$-ghost in this limit.

\subsection{Brane decay in the presence of other branes}

In this subsection, we show that one can extend our cohomology
arguments to the case where we include other branes that have not
decayed.  Consider OSFT around a 2 brane background, which we describe
by adding Chan-Paton indices to our string fields;
\begin{equation}
  \phi =
\left(\begin{matrix}
  \Phi_{11} & \Phi_{12}
\\
  \Phi_{21} & \Phi_{22}
\end{matrix}
\right),
\end{equation}
where $\phi^\dagger = \phi$.  To decay one of the branes, we may turn on
\begin{equation}
  \psi =
\left(\begin{matrix}
  \Psi & 0
\\
  0 & 0
\end{matrix}
\right).
\end{equation}
The BRST operator $Q_\psi$ acts as
\begin{equation} \label{multiBRST}
  Q_B \phi + [\psi ,\phi]
=
  \left(\begin{matrix}
  Q_B \Phi_{11} + [\Psi,\Phi_{11}]  & Q_B \Phi_{12} + \Psi \starp \Phi_{12}
\\
  Q_B \Phi_{21} -
    (-1)^{\text{gh}(\Phi_{21})} \Phi_{21} \starp \Psi  & Q_B \Phi_{22}
\end{matrix}
\right).
\end{equation}
Since we have decayed the first brane, we expect that there are no
on-shell $11$, $12$ or $21$ strings.  This implies that the three
BRST-operators,
\begin{equation}
  Q_{11}\Phi = Q_B\Phi + [\Psi , \Phi], \qquad
 Q_{12}\Phi = Q_B\Phi + \Psi\starp\Phi ,
\qquad \text{and} \qquad
 Q_{21}\Phi = Q_B\Phi - (-1)^{\text{gh}(\Phi)}\Phi \starp \Psi,
\end{equation}
should all have vanishing cohomology.  Since $Q_{11} = Q_\Psi$, there
is nothing new to show.  For $Q_{12}$ and $Q_{21}$, our old argument
still works as long as we are careful about left multiplication versus
right multiplication.  Suppose that $Q_{12} \Phi = 0$. Then it is easy
to check that
\begin{equation}
  Q_{12} (A\starp \Phi) =(Q_{\Psi} A) \starp \Phi = \Phi.
\end{equation}
Thus, as we expect, every closed state is exact.
Similarly, if $Q_{21} \Phi = 0$, we have
\begin{equation}
  Q_{21} (-\Phi \starp A) = \Phi \starp (Q_\Psi A) = \Phi.
\end{equation}
Putting the $A$ on the left of $\Phi$ would not work.  Hence, we have
shown that the only open strings that remain in the spectrum are those
that live on the undecayed brane.  This argument generalizes to the
case of $n$ decayed branes and $m$ undecayed branes in the expected
way.

\sectiono{Pure-gauge-like form}
\label{s:pg}
One of the curious features of the analytic tachyon vacuum is that it
is very close to being pure gauge.  Indeed, it was found by Okawa
\cite{Okawa:2006vm} that if one ignores the $\psi_N$ term -- which one
can in the $L_0$ basis\footnote{Interestingly, in the $\ll_0$ level
truncation we find $\Psi_\lambda = \frac{\lambda}{1-\lambda} Q\Phi +
\cdots$, where the dots stand for terms of $\ll_0$-level higher than
$0$. Hence, the $\lambda \to 1$ limit does not exist in this basis.}
-- the full solution can be written as the limit, $\lambda \to 1$, of
the state,
\begin{equation}
  \Psi_\lambda = U_\lambda \starp Q_B V_\lambda,
\end{equation}
where\footnote{$V_\lambda$ is defined by the Taylor series,
\begin{equation}
  \frac{1}{1-\lambda \Phi} = \sum_{n=0}^\infty \lambda^n \Phi^n;
\qquad \Phi^n = \underbrace{\Phi \starp \Phi \starp \ldots \starp \Phi}_{n}.
\end{equation}}
\begin{equation}
  U_\lambda = 1-\lambda \Phi, \qquad V_\lambda = \frac{1}{1-\lambda \Phi}
\end{equation}
and
\begin{equation}
  \Phi = B_1^L c_1 |0\rangle.
\end{equation}
When $\lambda <1$ the states $U_\lambda$ and $V_\lambda$ are well
defined in the level-expansion and the state $\Psi_\lambda$ is a true
pure-gauge solution with zero energy.

Obviously, the tachyon solution itself, cannot be a pure-gauge
solution related by a continuous deformation to the vacuum for two
reasons. First, the energy of such a solution would have to be zero in
contradiction with the now proven Sen's first conjecture. Second, it
would imply that the cohomology of the kinetic operator at the true
vacuum would be isomorphic to the cohomology of $Q_B$ in contradiction
with Sen's third conjecture.  It is therefore interesting to
understand how the solution ceases to be a pure gauge at $\lambda=1$
and how Sen's conjectures are rescued.

The basic property of the pure-gauge solutions is that
\begin{equation}\label{VUandUV}
  U_\lambda \starp V_\lambda = V_\lambda \starp U_\lambda = \mathcal{I}.
\end{equation}
This allows one to define an isomorphism between the states in the
perturbative vacuum and their corresponding states in the pure-gauge
vacuum;
\begin{equation}
  \phi \to \mathcal{F}_\lambda[\phi]= U_\lambda \starp \phi \starp V_\lambda,
\end{equation}
which has inverse, $\mathcal{F}^{-1}_\lambda[\phi] = V_\lambda \starp \phi \starp U_\lambda$.

This isomorphism relates the original BRST operator, $Q_B$, to the new BRST operator, $Q_\lambda$,
around the pure-gauge vacuum;
\begin{equation}\label{BRSTisomorphism}
   Q_\lambda (\mathcal{F}_\lambda[\phi])= \mathcal{F}_\lambda [Q_B \phi].
\end{equation}
It follows that the two operators have identical cohomology.

We can now ask how (\ref{VUandUV})-(\ref{BRSTisomorphism}) break down
when $\lambda \to 1$.  Clearly, since the right hand side of
(\ref{VUandUV}) is independent of $\lambda$, we will find
$\lim_{\lambda \to 1} U_\lambda \starp V_\lambda = \lim_{\lambda \to
1} V_\lambda \starp U_\lambda = \mathcal{I}$.  However, the state
$V_\lambda$ by itself diverges in the $\ll_0$ level-expansion,
although it appears to remain finite in the $L_0$ expansion.

Similar divergences occur when we consider $\mathcal{F}_{\lambda}
(\phi)$ and its inverse.  For concreteness, take $\phi = c \mathcal{O}
|0\rangle$, where $\mathcal{O}$ is a matter operator that satisfies
\begin{equation}
  [\ll_0,\mathcal{O}] = h \mathcal{O}.
\end{equation}
Following the rules of \cite{Schnabl:2005gv} we find
\begin{multline}
\mathcal{F}_\lambda[\phi] = c\mathcal{O}(0)\ket{0} + \sum_{m=1}^\infty \lambda^m U_{m+2}^\bpz
U_{m+2} \left\lbrace \frac{1}{2}  \tilde{\mathcal{O}}(x) (\tilde c(x)+ \tilde c(-x)) + \frac{1}{2}
\tilde{\mathcal{O}}(y) (\tilde c(x) - \tilde c(y)) \right.
\\
\left. - \frac{1}{\pi} (\mathcal{B}_0+\mathcal{B}^\bpz) \left( \tilde{\mathcal{O}}(x)\tilde
c(x)\tilde c(-x)+ \tilde{\mathcal{O}}(y)
 (\tilde c(x) - \tilde c(y))\tilde c(-x) \right)\right\rbrace \ket{0} ,
\end{multline}
where, for brevity, we have introduced $x=\frac{\pi}{4} m$ and
$y=\frac{\pi}{4} (m-2)$. Using this form, it is straightforward to work
out the coefficients in the $\ll_0$ basis
\begin{align}
\mathcal{F}_\lambda[\phi]  = & \frac{1}{1-\lambda} c\mathcal{O}(0) \ket{0} + \nonumber \\ &
+\frac{\lambda}{(1-\lambda)^2} \left[ -\frac{1}{2} \left(\ll_0 + \ll_0^\bpz\right) \tilde
c\tilde{\mathcal{O}}(0) + \left(\bb_0 + \bb_0^\bpz\right) \tilde c\tilde\partial \tilde c
\tilde{\mathcal{O}}(0) + \frac{\pi}{4} \bigl((1-\lambda)\tilde\partial \tilde c
\tilde{\mathcal{O}}(0) + \tilde c \tilde\partial \tilde{\mathcal{O}}(0)\bigr) \right] \ket{0} +
\nonumber\\ & + \cdots,
\end{align}
where the dots stand for terms of higher $\ll_0$-level.  We see that,
due to the presence of poles at $\lambda=1$, the state
$\mathcal{F}_{\lambda=1}[\phi]$ does not make sense in this
basis. Note that one cannot rescale $\phi_\lambda$ by a positive power
of $1-\lambda$ to get a finite representative of the cohomology, since
the maximal order of the poles grows with level.

We find similar behavior when we compute $\mathcal{F}[\phi]$ in the
ordinary $L_0$ level truncation.  Since such computations are more
difficult, we have restricted ourselves to the case where
$\mathcal{O}$ is a weight one primary.  This case is of particular
interest, as any cohomology class of $Q_B$ has a representative of
this form.

Computing the coefficient of  $\mathcal{F}_\lambda[\phi]$ in front of $c\mathcal{O}(0)\ket{0}$,  we find
\be
1+\sum_{m=1}^\infty \lambda^m \frac{m+2}{2} \left[\frac{1}{2} +
\frac{\alpha}{\pi}\left(1-\left(\frac{\sin\alpha}{\sin
2\alpha}\right)^2\right) - \frac{1}{\pi} \sin 2\alpha + \frac{1}{2\pi}
\left(\frac{\sin\alpha}{\sin 2\alpha}\right)^2 \left(\sin 4\alpha
-\sin 2\alpha \right) \right], 
\ee 
where $\alpha = \pi/(m+2)$. Since
the summand behaves as \bdm \lambda^m \left[ \frac{m+2}{4}
-\frac{1}{2} + \frac{\pi^2}{12} \frac{1}{(m+2)^2} + \cdots \right],
\edm
we see that, apart from the mild polylogarithmic singularities at
$\lambda=1$, which are present also for the solution $\Psi_\lambda$
itself, $\mathcal{F}_\lambda[\phi]$ contains double and single poles
and therefore the limit $\lim_{\lambda \to 1}
\mathcal{F}_\lambda[\phi]$ does not exist.

So far in this discussion we have tried to show that elements of the
cohomology of $Q_B$ are not mapped via $\mathcal{F}$ to elements of
the cohomology of $Q_\Psi$.  However, it is also interesting to ask
why $A$ cannot be pulled back to the perturbative vacuum to show that
$Q_B$ has no cohomology.  Hence, we compute
\begin{multline}
  \mathcal{F}^{-1} (A)
   = \frac{1}{1-\Phi} \starp A \starp (1-\Phi)
    = \tfrac{\pi}{2} \int_1^2 dr\, (1+ \sum_{n=1}^\infty B_1^L |n\rangle \starp c_1 |0\rangle)
    \starp B_1^L |r\rangle \starp (1-B_1^L c_1 |0\rangle)
    \\
    =  \tfrac{\pi}{2} \sum_{n=1}^\infty \int_1^2 dr\, B_1^L |n+r-1\rangle.
\end{multline}
This simplifies to
\begin{equation}
  \mathcal{F}^{-1}(A) = \tfrac{\pi}{2} \int_1^\infty dr\, B_1^L |r\rangle,
\end{equation}
which should be thought of as the ``$A$" of the perturbative vacuum.
We can now act on this state with $Q_B$;
\begin{equation} \label{eq:QFA}
  Q_B(\mathcal{F}_{-1}(A)) = -\int_1^\infty dr\, \partial_r |r\rangle =
  \mathcal{I} - |\infty\rangle.
\end{equation}
Happily, we do not find just the identity on the right hand side, so
the cohomology of $Q_B$ need not vanish.\footnote{Formally one could
write $Q_B \mathcal{F}^{-1}(A)=\mathcal{F}^{-1}(Q_\Psi
A)=\mathcal{F}^{-1}(\mathcal{I}) = V_{\lambda=1} \starp
U_{\lambda=1}$. Using (\ref{eq:QFA}), this would imply $V \starp
U=\mathcal{I} - |\infty\rangle$ suggesting that $V$ and $U$ are a
nontrivial pair of partial isometries as first proposed in
\cite{Schnabl:2000cp}. On the other hand a direct computation seems to
yield $V \starp U=\mathcal{I}$ in the strict $\lambda \to 1$ limit, in
both $L_0$ and $\mathcal{L}_0$ level truncation. It would be nice to
understand this anomaly more deeply.} Equation (\ref{eq:QFA}) has a
nice interpretation in terms of half strings. Consider a state $\phi$
which is $Q_B$-closed, but whose left half has no overlap with the
right half of $|\infty\rangle$. In other words, $|\infty\rangle \starp
\phi = 0$. It follows that $\phi$ is $Q_B$-exact.  To see this,
consider
\begin{equation}
  Q_B(\mathcal{F}^{-1}(A) \starp \phi ) = (\mathcal{I} -|\infty\rangle ) \starp\phi = \phi.
\end{equation}
A similar result holds for states whose right half has no overlap with the left half of
$|\infty\rangle$.  This implies that the entire cohomology of $Q_B$ should be found on states whose
left and right halves are given by the left and right halves of $|\infty\rangle$. Such a set of
states is easy to find.  For example, at ghost number $0$, the cohomology of $Q_B$ is represented
by just $|\infty\rangle$ itself.  At ghost number 1, which is the interesting case, the cohomology
of $Q_B$ has representatives given by weight $(0,0)$ primaries of the form $c J$, where $J$ is a
weight one matter primary. Inserting these operators at the midpoint of $|\infty\rangle$ gives a
set of ghost number $1$ states in the cohomology of $Q_B$ with left and right halves given by the
left and right halves of $|\infty \rangle$.

%
%
\section*{Acknowledgments} We would like to thank T.~Grimm, A.~Hashimoto, L.~Motl, Y.~Okawa, E.~Witten and B.~Zwiebach
for useful conversations.  I.E. is supported in part by the DOE grant DE-FG02-95ER40896 and by
funds from the University of Wisconsin.

\bibliography{c}\bibliographystyle{utphys}

\end{document}